\documentclass{ifacconf}

\usepackage{mathptmx} 
\usepackage{times} 

\usepackage{graphicx}      
\usepackage{natbib}        
\usepackage{amssymb,amsmath,amsfonts}

\newtheorem{definition}{Definition}

\def\endproof{\hfill{$\blacksquare$}}

\def\d{{\rm \partial}}        

\def\rd{{\rm d}}        
\def\re{{\rm e}}        
\def\bD{{\bf D}}
\def\bA{{\mathbf A}}
\def\bR{{\mathbf R}}
\def\bE{{\bf E}}
\def\cA{{\cal A}}
\def\fA{{\frak A}}
\def\cN{{\cal N}}
\def\cS{{\cal S}}

\def\bh{{\mathbf h}}

\def\mR{{\mathbb R}}    
\def\mL{{\mathbb L}}    

\def\mG{{\mathbb G}}    
\def\cov{{\bf cov}}    
\def\cH{{\mathcal H}}
\def\<{\leqslant}           
\def\>{\geqslant}           

\def\Tr{{\rm Tr}}
\def\rT{{\rm T}}
\def\sn{{| \! | \! |}}
\def\x{{\times}}

\def\mW{{\mathbb W}}    %

\begin{document}

\pagestyle{headings}
\begin{frontmatter}
\title{Anisotropic Norm Bounded Real Lemma for Linear Discrete Time Varying Systems\thanksref{footnoteinfo}}

\thanks[footnoteinfo]{EAM and APK are supported by the Russian Foundation for Basic
Research grant 08-08-00567. IGV is supported by the Australian
Research Council.\newline$^{\dagger}$ We deeply regret the untimely loss of Dr Eugene Maximov who passed away on 26 July 2010.  His talents and ideas will never be forgotten.
}
\author[First]{Eugene A. Maximov$^{\dagger}$}
\author[Second]{Alexander P. Kurdyukov}
\author[Third]{Igor G. Vladimirov}
\address[First]{Bauman Moscow State Technical University, Russia}
\address[Second]{Institute of Control Sciences, Russia (akurd@ipu.ru)}
\address[Third]{University of New South Wales at the Australian Defence Force Academy, Canberra, ACT 2600, Australia (igor.g.vladimirov@gmail.com)}

\begin{abstract}                
We consider a finite horizon linear discrete time varying system whose input is a random noise with an imprecisely known probability law. The statistical uncertainty is described by a nonnegative parameter $a$ which constrains the anisotropy of the noise as an entropy theoretic measure of deviation of the actual noise distribution from Gaussian white noise laws with scalar covariance matrices. The worst-case disturbance attenuation capabilities of the system with respect to the statistically uncertain random inputs are quantified by the $a$-anisotropic norm which is a constrained operator norm of the system. We establish an anisotropic norm bounded real lemma which provides a state-space criterion for the $a$-anisotropic norm of the system not to exceed a given threshold. The criterion is organized as an inequality on the determinants of matrices associated with a difference Riccati equation and extends the Bounded Real Lemma of the $\cH_{\infty}$-control theory. We also provide a necessary background on the anisotropy-based robust performance analysis.
\end{abstract}

%
\end{frontmatter}

\section{Introduction}
The statistical uncertainty, present in random disturbances as a discrepancy between the imprecisely known true probability distribution of the noise and its nominal model, may corrupt the expected performance of  a stochastic control system if the controller design is oriented at a specific probability law  of the disturbance. Such uncertainties result not only from the lack of prior knowledge of the actual noise statistics, but also from the inherent  variability of the environment where the control system operates.

The robustness in stochastic control can therefore be achieved by explicitly incorporating
different scenarios of the noise distribution into a single performance index to be optimized. The degree of robustness depends on the ``size'' of the uncertainty  used in the controller design. The statistical uncertainty can be measured in entropy theoretic terms and the robust performance index can be chosen so as to quantify the worst-case disturbance attenuation capabilities of the system.
It is this combination of approaches that underlies the anisotropy-based theory of stochastic robust control which was initiated about sixteen years ago at the interface of the entropy/information and robust control theories in a series of papers (\cite{SVK_1994,VKS_1995_1,VKS_1995_2,VKS_1996_1,VKS_1996_2,VKS_1999}). This theory employs the anisotropy functional as an entropy theoretic measure of deviation of the unknown actual noise distribution from the family of Gaussian white noise laws with scalar covariance matrices. Accordingly, the role of a robust performance index is played by the $a$-anisotropic norm $\sn F \sn_a$ of a system $F$ which is defined as the largest ratio of the root mean square (RMS) value of the output of the system to that of the input, provided that the anisotropy of the input disturbance does not exceed a given nonnegative parameter $a$. Thus, the input anisotropy level $a$ is the size of the statistical uncertainty, and the $a$-anisotropic norm of the system $\sn F\sn_a$ is the worst-case RMS gain which, in the framework of the disturbance  attenuation paradigm,   is to be minimized.

An important property of the $a$-anisotropic norm is that it coincides with a rescaled Frobenius (or $\cH_2$) norm of the system for $a=0$ and converges to the induced (or $\cH_{\infty}$) norm as $a\to+\infty$.  Therefore, $\sn \cdot \sn_a$ is an anisotropy-constrained stochastic version of the induced norm of the system which occupies a  unifying intermediate position between the $\cH_2$ and $\cH_{\infty}$-norms utilised as performance criteria in  the linear quadratic Gaussian (LQG) (\cite{KS_1972}) and $\cH_{\infty}$-control theories (\cite{DGKF_1989}).

In its original infinite horizon time invariant setting, the anisotropy-based theory employed  the anisotropy production rate per time step in a stationary  Gaussian random sequence. The mean anisotropy has useful links with the condition number of the covariance matrix and the transient time in the sequence, thus describing the amount of spatial non-roundness and temporal correlation in it. These connections have recently been revisited in (\cite{KV_2008}).

At the performance analysis level, the anisotropy-based theory was developed for time invariant systems in (\cite{VKS_1996_1}), where equations were obtained for computing the $a$-anisotropic norm in state space. An
extended exposition of that work can be found in (\cite{DVKS_2001}) and a generalization to finite horizon time varying systems  is provided by (\cite{VDK_2006}).

A state-space solution  to the anisotropy-based optimal control problem (which seeks an internally stabilizing controller to minimize  the $a$-anisotropic norm of the closed-loop system) was obtained in (\cite{VKS_1996_2}). The solution was found  as a saddle point in the stochastic minimax problem.
The anisotropy-based theory therefore offers tools both for the quantitative description of uncertainty and robust performance analysis on the one hand, and control design on the other.

The present paper is concerned with the anisotropy-based robust performance analysis of linear discrete time varying (LDTV) systems in state space. The procedure developed in (\cite{VDK_2006}) for computing the $a$-anisotropic norm of such a system over a finite time horizon involves the solution of three coupled equations: a backward difference Riccati equation, a forward difference Lyapunov equation, and an algebraic equation for the determinants of matrices associated with the previous equations.  Its practical implementation is complicated by the opposite time ordering of the coupled difference equations.

Here, we develop an alternative state-space criterion for the $a$-anisotropic norm to be bounded by a given threshold. This allows the above issue to be  overcome by eliminating the Lyapunov equation, replacing the backward Riccati equation by a forward Riccati equation,  and replacing the algebraic equation by an appropriately modified inequality. The resulting Anisotropic Norm Bounded Real Lemma (ANBRL) is organised as an inequality on matrices associated with a forward difference Riccati equation. In addition to a significant simplification of the previously developed anisotropy-based robust performance analysis (which can now be carried out recursively in time), ANBRL also provides an extension of the Bounded Real Lemma from the $\cH_{\infty}$-control theory to the uncertain stochastic setting with finite horizon time varying dynamics. An infinite-horizon version of ANBRL for time invariant systems, which involve algebraic equations free from the opposite time ordering issue, is presented in (\cite{KMT_2010}).

Another approach to robust control in stochastic systems,  using the relative entropy to describe statistical uncertainty,  can be found in (\cite{PJD_2000,P_2006,UP_2002}),   where an important role is played by a link between a relative entropy duality relation and robust properties of risk-sensitive controllers  (\cite{DJP_2000}) which minimize the expected-exponential-of-quadratic functional.
Although the ideas of entropy-constrained induced norms and associated stochastic minimax find further development in the control literature   (\cite{CR_2007}), the anisotropy-based theory of stochastic robust control remains largely unnoticed.  It is partly for this reason that the main result of the present paper
is preceded by the background material to assist the readers to build awareness of the anisotropy-based approach.

The paper is organised as follows. Section~\ref{sec:systems} specifies the class of systems being considered. Sections~\ref{sec:anisotropy} and \ref{sec:norm} provide the necessary background on the anisotropy of random vectors and the $a$-anisotropic norm of matrices. Section~\ref{sec:ANBRL} establishes the Anisotropic Norm Bounded Real Lemma. Its connection with the Bounded Real Lemma in the limit $a\to +\infty$ is discussed in Section~\ref{sec:Hinf_limit}. Section~\ref{sec:conclusion} gives concluding remarks. Appendix provides a subsidiary state space criterion of outerness.

\section{Class of systems being considered}
\label{sec:systems}

We consider a linear discrete time varying (LDTV) system $F$ on a
bounded time interval $[0,N]$. Its $n$-dimensional state $x_k$ and
$r$-dimensional output $z_k$ at time $k$ are governed by the
equations
\begin{eqnarray}
    \label{realF_x}
        x_{k+1}
        & = &
        A_k x_k + B_k w_k, \\
    \label{realF_z}
        z_k
        & = &
        C_k x_k + D_k w_k,
\end{eqnarray}
with initial condition $x_0=0$, which are driven by an 
$m$-dimensional input $w_k$. Here, $A_k$, $B_k$, $C_k$, $D_k$ are
appropriately dimensioned real matrices which are assumed to be known
functions of time $k$. The state-space equations
(\ref{realF_x})--(\ref{realF_z}) are written as
\begin{equation}
\label{realF}
    F =
        \begin{matrix}
         &  \!\!\!{}_{\leftarrow n\rightarrow\leftarrow m\rightarrow}\\
            \begin{matrix}
                {}^n & \updownarrow \\
                {}_r & \updownarrow
            \end{matrix} &
            \hskip-1.5mm
            \left[
                \begin{array}{c|c}
                    \,A_\bullet\, & \,B_\bullet\, \\
                    \hline
                    \,C_\bullet\, & \,D_\bullet\, \\
                \end{array}
            \right]\\
            {}
        \end{matrix},
\end{equation}
where we have also shown the dimensions. For any two moments of time
$s \< t$, the values of the input and output signals $W$ and $Z$ on
the interval $[s,t]$ are assembled into the column-vectors
$$
W_{s:t} :=
                [
                  w_s^{\rT}, \ldots
                  w_t^{\rT}
                ]^{\rT},
                \qquad
    Z_{s:t} :=
                [
                  z_s^{\rT}, \ldots,
                  z_t^{\rT}
                ]^{\rT}.
$$
Since the state of the system is zero-initialized, then
$Z_{0:t}=F_{0:t}W_{0:t}$, where $F_{s:t}$ is a block lower triangular matrix with $(r \x m)$-blocks $f_{jk}$ given by
\begin{equation}
\label{Fst}
    F_{s:t} := {\rm block}_{s \< j,k \< t} (f_{jk}),
    \ \ \
    f_{jk} = \left\{\begin{matrix}
                      C_j \Phi_{j,k+1} B_k     & {\rm for}\  j>k \\
                      D_k                   & {\rm for}\  j=k \\
                      0                     & {\rm otherwise}
                    \end{matrix}
                    \right..
\end{equation}
Here,
\begin{equation}
\label{Tjk}
    \Phi_{jk} := A_{j-1} \x \ldots \x A_k
\end{equation}
is the state transition matrix from $x_k$ to $x_j$ for $j\> k$, with
$\Phi_{kk} = I_n$ the identity matrix of order $n$. Since the
matrix $F_{0:N}$ completely specifies the system $F$ on the time
interval $[0,N]$ as a linear input-output operator from $W_{0:N}$ to
$Z_{0:N}$, all the norms of $F$ are those of $F_{0:N}$. In
particular, the finite-horizon counterparts of the $\cH_2$ and
$\cH_{\infty}$-norms are described by the Frobenius and operator
norms of $F_{0:N}$ as
\begin{equation}
\label{H2_Hinf}
    \|F\|_2
    :=
    \sqrt{
        \Tr
        (
            F_{0:N}^{\rT}
            F_{0:N}
        )
    },
    \qquad
    \|F\|_{\infty}
    :=
    \sigma_{\max}(F_{0:N}),
\end{equation}
where $\sigma_{\max}(\cdot)$ is the largest singular value of a
matrix. We will be concerned with the $a$-anisotropic norm of the
system $F$ which is also understood in terms of the matrix
$F_{0:N}$. This norm is obtained by modifying the concept of induced
norm with the aid of an additional constraint on the input which
involves the entropy theoretic construct of anisotropy.

\section{Anisotropy of random vectors}
\label{sec:anisotropy}

The relative entropy (or Kullback-Leibler informational divergence) (\cite{CT_2006})
 of a probability measure $P$ with respect to
another probability measure $M$ on the same measurable space is
defined by
$$
        \bD(P \| M)
        :=
                \bE
                \ln
                \frac{\rd P}{\rd M}.
$$
Here, $P$ is assumed to be absolutely continuous
with respect to $M$ with density (Radon-Nikodym derivative) $\rd
P /\rd M$, and $\bE$ denotes the expectation in the sense of $P$.
The relative entropy $\bD(P\|M)$, which  is always nonnegative,  
vanishes only if $P = M$.

In what follows, $\bD(P\| M)$ will also be written as $\bD(\xi\|
\eta)$ or $\bD(f\| g)$ if the probability measures $P$ and $M$ are
distributions of random vectors $\xi$ and $\eta$ or are specified by
their probability density functions (PDFs) $f$ and $g$ with respect
to a common measure.

For any $\lambda > 0$, we denote by $p_{\ell,\lambda}$ the
$\ell$-variate Gaussian PDF with zero mean and scalar covariance
matrix $\lambda I_{\ell}$:
\begin{equation}
    \label{pll}
        p_{\ell,\lambda}(w)
        =
        (2\pi \lambda)^{-\ell/2}
        \re^{-|w|^2/(2\lambda)},
        \qquad
        w \in \mR^{\ell}.
\end{equation}
Let $W$ be a square integrable absolutely continuous  random vector
with values in $\mR^{\ell}$ and PDF $f$. Its relative entropy with
respect to the Gaussian probability law (\ref{pll}) is computed as
\begin{eqnarray}
\nonumber
        \bD
        (
            f \| p_{\ell,\lambda}
        )
          & =&
        \bE \ln \frac{f(W)}{p_{\ell,\lambda}(W)}\\
\label{Dpll}
         & = &
        \frac{\ell}{2} \ln (2\pi \lambda)
        +
        \frac{\bE (|W|^2)}{2\lambda}
        -\bh(W),
\end{eqnarray}
where
$$
    \bh(W)
    :=
    -\bE \ln f(W)
    =
    -
    \int_{\mR^{\ell}}
    f(w) \ln f(w)
    \rd w
$$
is the differential entropy (\cite{CT_2006}) of $W$.
For what follows, the class of square integrable absolutely
continuous $\mR^{\ell}$-valued random vectors is denoted by
$\mL_2^{\ell}$.

\begin{definition} (\cite{VKS_1995_1,VDK_2006})
The aniso\-tropy $\bA(W)$ of a random vector $W \in \mL_2^{\ell}$ is
defined
 as the minimum relative entropy (\ref{Dpll}) of its PDF $f$ with respect to
 the Gaussian PDFs (\ref{pll}) with zero
mean and scalar covariance matrices:
\begin{equation}
\label{aniso}
        \bA(W)
         :=
         \inf_{\lambda > 0}
        \bD
        (
            f \| p_{\ell,\lambda}
        )
         =
        \frac{\ell}{2}
        \ln
        \frac{
                2\pi \re
                \bE (|W|^2)}
                {
                \ell
        }
        -
        \bh(W).
\end{equation}
\end{definition}

A similar construct to the rightmost expression in (\ref{aniso})
was considered for scalar random variables in a context of time
series prediction in \cite[Definition~4 on p.~2911]{B_1998}.

The minimum with respect to $\lambda$ in (\ref{aniso}) is achieved
at $\lambda = \bE(|W|^2)/\ell$. The corresponding ``nearest''
zero-mean Gaussian random vector $W_*$ has the covariance matrix
$\cov(W_*)= \bE(|W|^2)I_{\ell}/\ell$, and its differential entropy
coincides with the first term on the right-hand side of
(\ref{aniso}), that is,      $
        \bh(W_*)
        =
        \ell
        \ln
        (
            2\pi \re
            \bE (|W|^2)
            \big/
            \ell
        )/2
    $.
The class of $\mR^{\ell}$-valued Gaussian random vectors $W$ with
zero mean and a given nonsingular covariance matrix $\Sigma$ will be
written as $\mG^{\ell}(\Sigma)$ (it is a subclass of
$\mL_2^{\ell}$). Their PDF is
$$
    p(w)
    =
    (2\pi)^{-\ell/2}
    (\det \Sigma)^{-1/2}
    \re^{- \|w\|_{\Sigma^{-1}}^2/2}, 
$$
where  $\|v\|_M := \sqrt{\Tr(v^{\rT} M v)}$ is  the Euclidean
(semi-) norm of a vector $v$ weighted by a positive (semi-)
definite matrix $M$.

\begin{lem}(\cite{VKS_1995_1,VDK_2006})
\label{anisoprops}
\begin{itemize}
\item[{(a)}]
The anisotropy $\bA(W)$, defined by (\ref{aniso}), is invariant
under rotation and scaling of $W$, that is, $ \bA(\lambda U W) =
\bA(W) $ for any orthogonal matrix $U \in \mR^{\ell\x \ell}$ and any
$\lambda \in \mR\setminus \{0\}$;
\item[{(b)}]
The anisotropy of a random vector $W\in \mL_2^{\ell}$ with a given
matrix of second moments $\bE(WW^{\rT}) = \Sigma$ satisfies
$$
    \bA(W)
    \>
    -\frac{1}{2}
    \ln\det\frac{\ell\Sigma}{\Tr\Sigma}.
$$
This inequality holds as an equality if and only if $W$ is Gaussian
with zero mean and covariance matrix $\cov(W)=\Sigma$;

\item[{(c)}]
For any random vector $W \in \mL_2^{\ell}$, its anisotropy $\bA(W)$
is always nonnegative and vanishes only if $W$ is Gaussian
distributed with zero mean and scalar covariance matrix (that is,
$\cov(W) = \lambda I_{\ell}$ for some $\lambda
> 0$).
\end{itemize}
\end{lem}

Lemma~\ref{anisoprops}(a) shows that $\bA(W)$ quantifies the
rotational non-invariance of the PDF of $W$. This property
originally motivated the term ``anisotropy'' for the
functional. In application to Gaussian random vectors $W$, the
assertions (b) and (c) of the lemma allow $\bA(W)$ to be interpreted
as a measure of heteroscedasticity and cross-correlation of the
entries of $W$.

Furthermore, Lemma~\ref{anisoprops}(b) implies that if an
arbitrary random vector $W \in \mL_2^{\ell}$,  with second-moment
matrix $\Sigma:=\bE(WW^{\rT})$,  is replaced by a Gaussian vector
$\Gamma$ with zero mean $\bE(\Gamma) = 0$ and covariance matrix
$\cov(\Gamma)=\Sigma$, then the transition $W\mapsto \Gamma$ is an
anisotropy-decreasing operation which preserves the second-moment
matrix, that is,  $\bA(\Gamma) \< \bA(W)$ and $\bE(\Gamma\Gamma^{\rT}) =
\Sigma$.

Another important property of the anisotropy functional is
its superadditivity 
$$
    \bA
    (W)
    \>
    \bA(W_1) + \bA(W_2),
    \qquad
    W:= [W_1^{\rT}, W_2^{\rT}]^{\rT}
$$
with respect to partitioning the random vector
$W$ into subvectors $W_1$ and $W_2$ \cite[Lemma~3 on p.~1269]{VDK_2006}.
This superadditivity is closely related to the asymptotically linear
growth of the anisotropy for long segments of a stationary random
sequence that allows the mean anisotropy
 to be defined as the anisotropy production rate
per time step  (\cite{VKS_1995_1}).

\section{$\lowercase{a}$-anisotropic norm of matrices}
\label{sec:norm}

Let $F\in \mR^{s\x \ell}$ be an arbitrary matrix. We will interpret
it as a deterministic linear operator whose input is a square
integrable $\mR^{\ell}$-valued random vector $W$ which is considered
to be a disturbance. While the disturbance attenuation paradigm
seeks to minimize the magnitude of the output $Z:= FW$, the
probability distribution of $W$ can be regarded as the strategy of a
hypothetical player aiming to maximize the root-mean-square (RMS)
gain of $F$ with respect to $W$:
\begin{equation}
\label{RMS}
    \bR(F,W)
    =
    \sqrt{
    \frac{\bE(|Z|^2)}{
    \bE(|W|^2)}
    }.
\end{equation}
Here,  the squared Euclidean norm $|\cdot|^2$ of a vector    is
interpreted as its ``energy'', so that $\bE(|W|^2)$ and $\bE(|Z|^2)$
describe the average energy (or power) of  the input and output of
the operator $F$, respectively. The denominator $\bE(|W|^2)$ in
(\ref{RMS}) vanishes only  in the trivial case, where $W= 0$ with probability one, which is
excluded from consideration.

The map $F\mapsto \bR(F,W)$ is a semi-norm in $\mR^{s \x \ell}$. It
is a norm if and only if the matrix of second moments $\Sigma :=
\bE(WW^{\rT})$  of the random vector $W$ is nonsingular. Indeed,
since $\bE(|W|^2) = \Tr \Sigma$ and $\bE(|Z|^2) = \Tr (F \Sigma
F^{\rT})$, the semi-norm properties follow from the representation
of the RMS gain (\ref{RMS}) in terms of the Frobenius  norm
$\|\cdot\|_2$ as
$
    \bR(F,W)= \|F\Omega\|_2$, where
$
    \Omega :=\sqrt{\Sigma/\Tr\Sigma}
$.
This also shows that $\bR(F,W)=0$ implies
$F=0$ if and only if $\Omega$ is positive definite which is
equivalent to the positive definiteness of $\Sigma$.

The RMS gain $\bR(F,W)$ depends on the matrix $F$ only through
$F^{\rT}F$ and never exceeds the induced operator norm
$\|F\|_{\infty}$. If there are no restrictions on the probability
distribution of $W$ other than the square integrability $\bE(|W|^2)<
+\infty$, then $\bR(F,W)$ can be made arbitrarily close to its upper
bound $\|F\|_{\infty}$. This is achieved by concentrating the
distribution of $W$ along the eigen-space of the matrix $F^{\rT}F$
in $\mR^{\ell}$  associated with its largest eigenvalue
$\|F\|_{\infty}^2$. However, except when the matrix $F^{\rT}F$ is
scalar, such distributions are singular with respect to the
$\ell$-dimensional Lebesgue measure and should be considered as 
``non-generic''.

For any random vector $W \in \mL_2^{\ell}$, we quantify
``non-genericity'' of its probability distribution by the anisotropy
$\bA(W)$. Accordingly, we  assume that the disturbance player is
constrained by the condition $\bA(W) \< a$, where $a$ is a given
nonnegative parameter. In particular, if $a=0$, then by
Lemma~\ref{anisoprops}(c), the player is allowed to generate only 
Gaussian random vectors  $W\in\bigcup_{\lambda>0}\mG^{\ell}(\lambda
I_{\ell})$ with zero mean and scalar covariance matrices. With
respect to any such $W$, the RMS gain (\ref{RMS}) of the operator
$F$ becomes a scaled Frobenius norm:
$\bR(F,W)=\|F\|_2/\sqrt{\ell}$.

\begin{definition}
For any $a\> 0$, the $a$-anisotropic norm of a matrix $F\in \mR^{s\x
\ell}$ is defined as an anisotropy-constrained upper envelope of the
RMS gains (\ref{RMS}):
\begin{equation}
\label{aniso_norm}
    \sn F \sn_a
    :=
    \sup
    \{
        \bR(F,W) : \
        W\in\mL_2^{\ell}, \
        \bA(W) \< a
    \}.
\end{equation}
\end{definition}

This definition closely follows the concept of an induced norm, with
the only, though essential, difference being the constraint $\bA(W)
\< a$ on the anisotropy (\ref{aniso}). It is the latter point where
the entropy theoretic considerations enter the construct of the
$a$-anisotropic norm (\ref{aniso_norm}), thus making $\sn F\sn_a$ an
anisotropy-constrained stochastic version of the induced operator
norm $\|F\|_{\infty}$.

For any given matrix $F\in\mR^{s \x \ell}$,  the $a$-anisotropic
norm $\sn F \sn_a$  is a nondecreasing concave function of $a\> 0$, 
which satisfies
\begin{equation}
    \label{H2_Hinf_limits}
    \frac{\|F\|_2}{\sqrt{\ell}}
    =
    \sn F \sn_0
    \< \sn F \sn_a
    \<
    \lim_{a\to +\infty}
    \sn F \sn_a
    =
    \|F\|_{\infty}.
\end{equation}
The rate of convergence of $\sn F\sn_a$ to the limiting values
$\|F\|_2/\sqrt{\ell}$ and $\|F\|_{\infty}$ is investigated in
(\cite{VKS_1999,VDK_2006}). The relations (\ref{H2_Hinf_limits})
show that the $a$-anisotropic norm occupies an intermediate unifying
position between the  scaled Frobenius norm and the induced operator
norm.

Note that $b\|F\|_2/\sqrt{\ell} + (1-b) \|F\|_{\infty}$, with $b:=
\exp(-a)$, also provides an ``intermediate norm''. However, unlike
the naive convex combination of the extreme norms,  $\sn \cdot \sn_a$ is an anisotropy-constrained
operator norm, the very definition (\ref{aniso_norm}) of which is
concerned with the worst-case disturbance attenuation capabilities
of the linear operator (measured by the RMS gain (\ref{RMS})) with
respect to statistically uncertain random inputs (with the
uncertainty being measured by the anisotropy
 (\ref{aniso})) and combines both power and entropy concepts in a
 physically sound manner.

\section{Anisotropic norm bounded real lemma}
\label{sec:ANBRL}

In application to the LDTV system $F$ of Section~\ref{sec:systems},
the $a$-anisotropic norm $\sn F\sn_a:=\sn F_{0:N}\sn_a$, computed
for the matrix (\ref{Fst}), provides a robust performance index of
the system with respect to statistically uncertain random
disturbances $W$ over the time  interval  $[0,N]$. In this case, the
relations (\ref{H2_Hinf_limits}),  with $\ell := m(N+1)$ the
dimension of $W_{0:N}$, take the form
\begin{equation}
\label{H2_Hinf_limits_system}
    \frac{\|F\|_2}{\sqrt{m(N+1)}}
    =
    \sn F\sn_0
    \<
    \sn F\sn_a
    \<
    \lim_{a\to +\infty}
    \sn F\sn_a
    =
    \|F\|_{\infty},
\end{equation}
where the norms (\ref{H2_Hinf}) are used. For a time invariant system $F$,
its $a$-anisotropic norm on the  interval
$[0,N]$ with $a:=\alpha N$ tends to the $\alpha$-anisotropic norm of the
system as $N\to+\infty$. If the anisotropy level $a$ grows sublinearly
($a=o(N)$) or superlinearly ($a=1/o(1/N)$) with the time
horizon $N$, then $\sn F\sn_a$ converges to either of the
extreme norms of the time invariant system  $\|F\|_2/\sqrt{m}$ or $\|F\|_{\infty}$, respectively.

In (\cite{VDK_2006}),  computing the anisotropic norm of a finite horizon LDTV system in state space was reduced  to solving three coupled equations: a backward difference Riccati equation, an algebraic equation involving the determinants of matrices,  and a forward difference Lyapunov equation. The procedure of the anisotropy-based robust  performance analysis is complicated by the presence of coupled difference equations with opposite time ordering.

The theorem below provides a state-space criterion for the $a$-anisotropic norm to be bounded by a given threshold $\gamma$. It turns out that the above issue can be overcome by eliminating the Lyapunov equation and replacing the algebraic equation by an appropriately modified inequality. Moreover, the backward Riccati equation can be replaced by a forward Riccati equation. By analogy with the Bounded Real Lemma in the $\cH_{\infty}$-control theory, we call the theorem Anisotropic Norm Bounded Real Lemma.

\begin{thm}
\label{th:ANBRL} Let $F$ be an LDTV system with the state-space
realization (\ref{realF}).  Then its $a$-anisotropic norm on the
time interval $[0,N]$ satisfies $\sn F \sn_a \< \gamma$ if and only
if there exists $q\> 0$ such that for the matrices $R_k\in \mR^{n\x
n}$, with $k=0, \ldots, N$, governed by the difference
Riccati equation
\begin{eqnarray}
\label{RicR}
    R_{k+1}
    & = &
    A_k R_k A_k^{\rT}
    +
      q B_k B_k^{\rT}
     +
    M_k S_k M_k^{\rT}, \\
\label{RicM}
    M_k
    & = &
    -(
        A_k R_k C_k^{\rT}
        +
        q B_k D_k^{\rT}
    )    S_k^{-1},\\
\label{RicS}
    S_k
    & = &
        I_r - C_k R_k C_k^{\rT} - q D_k D_k^{\rT},
\end{eqnarray}
with the initial condition $R_0 = 0$,   the matrices $S_0,
\ldots, S_N$ are all positive definite and satisfy the inequality
\begin{equation}
        \label{special}
        \sum_{k=0}^{N}
        \ln\det S_k
        \>
        m(N+1)\ln(1-q\gamma^2)+2a.
    \end{equation}
\end{thm}
Prior to proving the theorem, note that the
matrices $S_0, \ldots, S_N$, defined by (\ref{RicS}), are all positive
definite if and only if $q< \|F\|_{\infty}^{-2}$. For any such $q$,
the left-hand side of (\ref{special}) is nonpositive, since $S_k
\preccurlyeq I_m$ (and so, $\ln\det S_k \< 0$). Hence, any $q$
satisfying the specifications of Theorem~\ref{th:ANBRL} must also
satisfy the inequalities
\begin{equation}
\label{q_range}
    \gamma^{-2}
    (
        1-
        \re^{-2\alpha/m}
    )
    \<
    q
    <
    \gamma^{-2},
    \qquad
    \alpha:= \frac{a}{N+1}.
\end{equation}
Here, the ratio $\alpha$ is the anisotropy production rate per time
step. Therefore, if $\alpha$ significantly exceeds the dimension $m$
of the input $W$, then (\ref{q_range}) yields a relatively narrow
localization of the candidate values for $q$ about  $\gamma^{-2}$.
\begin{pf}Consider a class $    \mW_a
    :=
    \{
        W_{0:N}\in\mL_2^{\ell} : \,
        \bA(W_{0:N}) \< a
    \}
$
 of square integrable absolutely continuous random inputs to the system $F$  on the time interval $[0,N]$ with the anisotropy (\ref{aniso}) bounded by $a$, where $
    \ell := m(N+1)$.
By applying (\ref{RMS})
and (\ref{aniso_norm}) to the matrix $F_{0:N}$ in (\ref{Fst}), with
which we identify the finite horizon system $F$, it follows that the
inequality $\sn F \sn_a \< \gamma$ is equivalent to the fulfillment of
$    \bR(F_{0:N},W_{0:N})
    =
    \sqrt{\Tr(\Lambda \Pi)}\< \gamma
$
  for all $
    W_{0:N} \in \mW_a
$.
Here, the RMS gain is completely specified by the matrices
\begin{equation}
\label{Lambda_Pi}
    \Lambda
    :=
    F_{0:N}^{\rT} F_{0:N},
    \qquad
    \Pi
    :=
    \frac{\Sigma}{\Tr\Sigma},
\end{equation}
with $
    \Sigma
    :=
    \bE(W_{0:N}W_{0:N}^{\rT})
$,
so that $\Pi$ can be any positive definite matrix of order $\ell$ with unit trace.
Application of Lemma~\ref{anisoprops}(b) to
$W_{0:N}$ yields the inequality
$    \bA(W_{0:N})
    \>
    -
    \ln\det(\ell \Pi)/2
$á
which becomes an equality if and only if $W_{0:N}$ is Gaussian distributed
with zero mean and covariance matrix $\cov(W_{0:N})=\sigma \Pi$ for some $\sigma >0$. Thus, the minimum anisotropy of the disturbance $W_{0:N}$, required to achieve a given value $\gamma$ of the RMS gain of the system, is
\begin{eqnarray}
\nonumber
    \min_{\bR(F,W_{0:N})\> \gamma} \bA(W_{0:N})
     & =&
    -
    \frac{1}{2}\,
    \max_{\Pi\succ 0:\, \Tr \Pi = 1, \Tr(\Lambda \Pi) \> \gamma^2}
        \ln\det(\ell \Pi)\\
\label{min_aniso}
    & =&
    \min_{0\< q< \|F\|_{\infty}^{-2}:\ \cN(q)\> \gamma}
        \cA(q),
\end{eqnarray}
and is delivered by zero mean Gaussian random vectors $W_{0:N}$ with covariance matrices proportional to
\begin{equation}
\label{S}
    \cS(q)
    :=
    (I_{\ell}-q\Lambda)^{-1},
\end{equation}
with $0\< q < \|F\|_{\infty}^{-2}$. Here, assuming that the matrix $\Lambda$ is not scalar (the  trivial case is excluded from consideration), the associated functions
\begin{eqnarray}
\label{A}
    \cA(q)
    &:=& -\frac{1}{2}\ln\det\frac{\ell \cS(q)}{\Tr \cS(q)},\\
\label{N}
    \cN(q)
    &:=&
    \sqrt{\frac{\Tr(\Lambda \cS(q))}{\Tr \cS(q)}}
\end{eqnarray}
are strictly increasing in $q$. This allows the $a$-anisotropic norm of the system to be computed as
$\sn F\sn_a = \cN(\cA^{-1}(a))$,
where $\cA^{-1}$ is the
functional inverse of $\cA$.  The solution of the constrained optimization problem (\ref{min_aniso}), described above, is obtained by using the method of Lagrange multipliers, the Frechet derivative $\d_{\Sigma} \ln\det \Sigma =\Sigma^{-1}$ and strict concavity of $\ln\det \Sigma$ on the cone of positive definite matrices $\Sigma$ \cite[Theorem~7.6.7 on p.~466]{HJ_2007}; see (\cite{DVKS_2001,VKS_1996_1,VDK_2006}) for details. However, instead of computing  the $a$-anisotropic norm $\sn F\sn_a$, we will proceed  directly  to considering the structure of its sublevel set $\sn F\sn_a\< \gamma$. To this end,
from (\ref{S}), it follows that
$    \Lambda
    = (I_{\ell}-\cS(q)^{-1})/q
$ and
\begin{equation}
\label{useful1}
    \Tr(\Lambda \cS(q)) = \frac{\Tr \cS(q)-\ell}{q},
\end{equation}
which, in combination with (\ref{N}), implies that
\begin{equation}
\label{useful2}
    \Tr \cS(q) = \frac{\ell}{1-q\cN(q)^2}.
\end{equation}
 Substitution of the last identity into (\ref{A}) yields
\begin{equation}
\label{AN}
    \cA(q)
    =
    \fA(q,\cN(q)),
\end{equation}
where
\begin{equation}
\label{fA}
    \fA(q,\gamma)
    :=
    \frac{1}{2}\ln\det(I_{\ell}-q\Lambda) -\frac{\ell}{2} \ln (1-q\gamma^2).
\end{equation}
Since $-\ln(1-q\gamma^2)$ is monotonically increasing in $\gamma\in [0,1/\sqrt{q})$, then so is $\fA(q,\gamma)$. In particular, by the strict monotonicity of the function $\cN$,
\begin{equation}
\label{beyond}
    \cA(q)
    \mathop{\<}^{(\>)}
    \fA(q,\gamma)
    \quad
    {\rm for}\
    q
    \mathop{\<}^{(\>)}
    \cN^{-1}(\gamma).
\end{equation}
Although the computation of the $a$-anisotropic norm $\sn F\sn_a$ requires both functions $\cA$ and $\cN$ from (\ref{A}) and (\ref{N}), the function $\fA(\cdot,\gamma)$ contains all the information about the system $F$ needed to decide whether it satisfies $\sn F \sn_a \< \gamma$. This is based on the property that the function $\fA(q,\gamma)$ achieves its maximum at
the point $q=\cN^{-1}(\gamma)$ where, in view of (\ref{AN}), it coincides with the function $\cA$:
\begin{equation}
\label{nota_bene}
    \max_{0\< q< \|F\|_{\infty}^{-2}}
    \fA(q,\gamma)
    =
    \fA(\cN^{-1}(\gamma),\gamma)
    =
    \cA(\cN^{-1}(\gamma)),
\end{equation}
as shown in Fig.~\ref{fig:check}. Before proving this property, we note that
\begin{figure}[htb]
\begin{center}
\includegraphics[width=6cm]{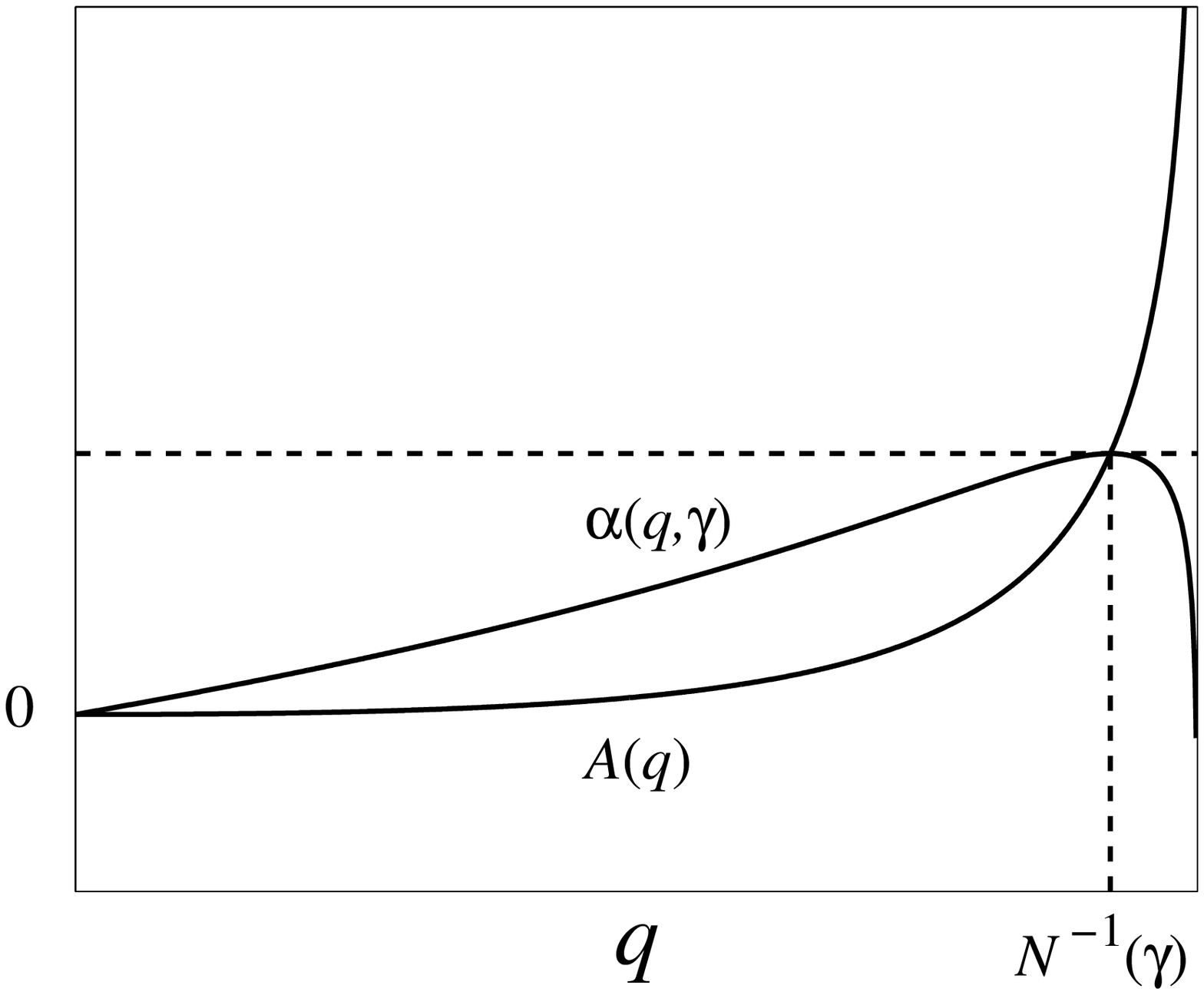}
\end{center}
\caption{Typical graphs of $\cA(q)$ and $\fA(q,\gamma)$ from (\ref{A}) and (\ref{fA}) as functions of $q$.
  The graphs intersect each other, and $\fA(q,\gamma)$ attains its maximum, at $q = \cN^{-1}(\gamma)$.
  The maximum value $\cA(\cN^{-1}(\gamma))$ is the minimum anisotropy of the input disturbance required to achieve the level $\gamma$ for the RMS gain of the system in (\ref{min_aniso}). The mutual position of the graphs for $q\ne \cN^{-1}(\gamma)$ is qualitatively described by (\ref{beyond}). }
  \label{fig:check}
\end{figure}
the inequality $\sn F\sn_a \< \gamma$ is equivalent to $\cA(\cN^{-1}(\gamma))\> a$. On the other hand, (\ref{nota_bene}) implies that $\cA(\cN^{-1}(\gamma))\> a$ is equivalent to the existence of $q\in [0, \|F\|_{\infty}^{-2})$ such that $\fA(q,\gamma)\> a$. Therefore,
\begin{equation}
\label{desired}
    \fA(q,\gamma)\> a\ 
    {\rm  for\ some}\ 
    q\ 
    \Longleftrightarrow\ 
    \sn F\sn_a\< \gamma.
\end{equation}
Now, to prove (\ref{nota_bene}), we differentiate the function $\fA$ from (\ref{fA}) in its first argument:
\begin{align}
\nonumber
    \d_q\fA(q,\gamma)
    =&
    \frac{1}{2}\d_q \ln \det(I_{\ell}-q\Lambda)
    +
    \frac{\ell \gamma^2}{2(1-q\gamma^2)}\\
\nonumber
    =&
    -\frac{1}{2}\Tr(\Lambda\cS(q))
    +
    \frac{\ell \gamma^2}{2(1-q\gamma^2)}\\
\nonumber
    & -\frac{\ell\cN(q)^2}{2(1-q\cN(q)^2)}
    +
    \frac{\ell \gamma^2}{2(1-q\gamma^2)}\\
\label{fAdiff}
    =&
    \frac{\ell(\gamma^2-\cN(q)^2)}{2(1-q\gamma^2)(1-q\cN(q)^2)},
\end{align}
where $\d_q(\cdot)$ is the partial derivative with respect to  $q$, and  use is made of (\ref{S}), (\ref{useful1}), (\ref{useful2}).
From (\ref{fAdiff}) and the strict monotonicity of $\cN$,  it follows that $\d_q\fA(q,\gamma)$ is positive for $q< \cN^{-1}(\gamma)$ and negative for $q > \cN^{-1}(\gamma)$, thereby establishing (\ref{nota_bene}).

We will now express the condition $\fA(q,\gamma)\> a$ on the function (\ref{fA}) in terms of the state-space dynamics of the system (\ref{realF_x})--(\ref{realF_z}) which has not been used yet. Recall that, for any
conformable matrices $M$ and $U$, the spectra of
$MU$ and $UM$  can differ from each other only by
zeros \cite[Theorem~1.3.20 on p.~53]{HJ_2007}. Hence, if we change
the order in which $F_{0:N}^{\rT}$ and $F_{0:N}$ are multiplied in the definition of
the matrix $\Lambda$ in (\ref{Lambda_Pi}), then it follows that the spectrum of $I_{\ell}-q\Lambda$
differs from that of $I_{r(N+1)}-qF_{0:N}F_{0:N}^{\rT}$ only by ones, and so
\begin{equation}
\label{dets}
    \det (I_{\ell} - q\Lambda)
    =
    \det(I_{r(N+1)}-qF_{0:N}F_{0:N}^{\rT}).
\end{equation}
Now, the latter matrix can be factorized for any $q< \|F\|_{\infty}^{-2}$ as
\begin{equation}
\label{FFHH}
    I_{r(N+1)} - qF_{0:N}F_{0:N}^{\rT}
    =
    H_{0:N}H_{0:N}^{\rT},
\end{equation}
where the matrix $H_{0:N}$ represents an LDTV system $H$, whose input and output are both $r$-dimensional, on the time interval $[0,N]$. The factorization (\ref{FFHH}) is equivalent to that an ancillary system
\begin{equation}
\label{Psi_def}
    \Psi
    :=
            \begin{bmatrix}
               \sqrt{q} F &
               H
            \end{bmatrix},
\end{equation}
with an $(m+r)$-dimensional input and $r$-dimensional output, is outer on the time interval $[0,N]$, that is, $\Psi_{0:N}\Psi_{0:N}^{\rT}= I_{r(N+1)}$. This property means that $\Psi$ transforms an $(m+r)$-dimensional Gaussian white noise sequence (with zero mean and identity covariance matrix) at the input into an  $r$-dimensional sequence with the same properties at the output. The system $H$ can be found in the form
\begin{equation}
\label{H}
    H
    =
    \left[
    \begin{array}{c|c}
      A_{\bullet} & M_\bullet \sqrt{S_\bullet}
      \\
      \hline
      C_\bullet & \sqrt{S_\bullet}
    \end{array}
    \right],
\end{equation}
where  the matrices $S_0, \ldots, S_N$ are all positive definite, so that, in view of (\ref{realF}),  the state-space realization of $\Psi$ in (\ref{Psi_def}) is
\begin{equation}
\label{Psi}
    \Psi
        =
        \left[
            \begin{array}{c|cc}
                A_\bullet & \sqrt{q}B_\bullet & M_{\bullet} \sqrt{S_{\bullet}}\\
                \hline
                 C_\bullet & \sqrt{q} D_\bullet & \sqrt{S_{\bullet}}
               \end{array}
             \right].
\end{equation}
Application of the state-space criterion of outerness from Appendix to (\ref{Psi}) yields the Riccati equation (\ref{RicR})--(\ref{RicS}), where  $R_0, \ldots, R_N$ are the controllability gramians of the system
$\Psi$. Since (\ref{H}) implies that $H_{0:N}$ is a block lower triangular matrix
, that is,
$$
    H_{0:N}
    =
        \begin{bmatrix}
            \sqrt{S_0} & & 0\\
             & \ddots\\
            * &  & \sqrt{S_N}
        \end{bmatrix}, 
$$ 
with the blocks $\sqrt{S_0}, \ldots, \sqrt{S_N}$ over the main diagonal, then (\ref{dets}) and (\ref{FFHH}) yield
$$
    \det (I_{\ell}-q\Lambda)
    =
    (\det H_{0:N})^2
    =
    \prod_{k=0}^{N}
    \det S_k.
$$
By substituting  this representation into (\ref{fA}), it follows that the inequality $\fA(q,\gamma)\> a$  in (\ref{desired}) is equivalent to (\ref{special}).
\endproof
\end{pf}

\section{Infinite anisotropy limit}
\label{sec:Hinf_limit}

If the anisotropy level increases unboundedly, $a\to +\infty$,   then the localization (\ref{q_range}), which follows from the inequality (\ref{special}), yields $q\to \gamma^{-2}$. In this case,  the Riccati equation (\ref{RicR})--(\ref{RicS}) takes the form
\begin{eqnarray}
\label{HinfRicR}
    R_{k+1}
    & = &
    A_k R_k A_k^{\rT}
    +
      \gamma^{-2} B_k B_k^{\rT}
     +
    M_k S_k M_k^{\rT}, \\
\label{HinfRicM}
    M_k
    & = &
    -(
        A_k R_k C_k^{\rT}
        +
        \gamma^{-2} B_k D_k^{\rT}
    )    S_k^{-1},\\
\label{HinfRicS}
    S_k
    & = &
        I_r - C_k R_k C_k^{\rT} - \gamma^{-2} D_k D_k^{\rT},
\end{eqnarray}
well-known in the context of $\cH_{\infty}$-suboptimal controllers.
This is closely related to the convergence $\lim_{a\to +\infty} \sn F\sn_a = \|F\|_{\infty}$ in (\ref{H2_Hinf_limits_system}), whereby the inequality $\sn F\sn_a \< \gamma$ ``approaches'' $\| F\|_{\infty} \< \gamma$ for large values of $a$. Therefore, in the limit, as $a\to +\infty$, Theorem~\ref{th:ANBRL} reduces to the Bounded Real Lemma, which establishes the equivalence between the inequality $\|F\|_{\infty}<\gamma$ and the positive definiteness of the matrices $S_0, \ldots, S_N$ associated with (\ref{HinfRicR})--(\ref{HinfRicS}).

\section{Conclusion}
\label{sec:conclusion}

We have considered a class of finite-dimensional linear discrete time varying systems on a bounded time interval subjected to input disturbances with an unknown probability law.

The statistical uncertainty has been quantified using the concept of anisotropy as an entropy theoretic measure of deviation of the actual noise distribution from nominal Gaussian white noise distributions with scalar covariance matrices.

The associated robust performance index, describing the worst-case disturbance attenuation capabilities of the system, is  the $a$-anisotropic norm defined as the largest root mean square gain of the system with respect to random noises whose anisotropy is bounded by a given nonnegative parameter $a$.

We have established a state-space criterion for the $a$-anisotropic norm not exceeding a given threshold value. The   Anisotropic Norm Bounded Real Lemma (ANBRL) is organised as an inequality on the determinants of matrices associated with a forward difference Riccati equation.

Apart from substantially  simplifying the previously developed procedure of anisotropy-based robust performance analysis, which is now amenable to a recursive implementation, ANBRL includes the Bounded Real Lemma of the $\cH_{\infty}$-control theory as a limiting case, thus extending it to the statistically uncertain stochastic setting with time varying dynamics.


\appendix*
\section{. State-space criterion of outerness}

In this section, we assume that the output dimension of the system $F$ in (\ref{realF}) does not exceed the input dimension: $r\< m$. Such a system is said to be \textit{outer} on the time interval $[0,N]$ if $F_{0:N} F_{0:N}^{\rT} = I_{r(N+1)}$. This is equivalent to the  preservation of the property of being a Gaussian white noise sequence (with zero mean and identity covariance matrix) for signals passing through the system. The state-space criterion of outerness, given below for completeness of exposition, is similar to that in the time invariant case (\cite{GTYP_1989}) and utilizes the controllability and observability gramians
\begin{equation}
\label{PQ_def}
    P_j
     :=
    \sum_{k=0}^{j-1}
    \Phi_{j,k+1}B_kB_k^{\rT} \Phi_{j,k+1}^{\rT},
\quad
    Q_k
     :=
    \sum_{j=k}^{N}
    \Phi_{jk}^{\rT} C_j^{\rT} C_j \Phi_{jk},
\end{equation}
where $\Phi_{jk}$ is the
state transition matrix (\ref{Tjk}). The gramians
satisfy the difference Lyapunov equations
$$    P_{j+1}
     =
    A_j P_j A_j^{\rT} + B_j B_j^{\rT},
    \qquad
    Q_k
     =
    A_k^{\rT} Q_{k+1} A_k + C_k^{\rT} C_k
$$
with initial condition $P_0=0$ and terminal condition $Q_{N+1}=0$.

\begin{lem}
\label{lem:outerness} The LDTV system $F$ with the state-space
realization (\ref{realF}) is outer on the time interval $[0,N]$ if
and only if for every $k=0,\ldots, N$,
\begin{equation}
\label{cond}
    C_k P_k C_k^{\rT} + D_k D_k^{\rT}
     =
    I_r,
    \qquad
    Q_{k+1}( A_k P_k C_k^{\rT} + B_k D_k^{\rT})
     =
    0.
\end{equation}
\end{lem}
\begin{pf}
Let the input $W$ to the system $F$ be an $m$-dimensional Gaussian white noise  sequence with zero mean and identity covariance matrix. Then, in view of (\ref{realF_z}),  the output $Z$ is also a zero mean Gaussian random sequence with
\begin{equation}
\label{covzk}
    \cov(z_k) = C_kP_k C_k^{\rT}+D_kD_k^{\rT},
\end{equation}
since $x_k$ and $w_k$ are independent, with $\cov(x_k)=P_k$ the $k$th controllability gramian from (\ref{PQ_def}). Consider the cross-covariance of $z_j$ and $z_k$ for $j>k\> 0$. By using the state transition matrix  (\ref{Tjk}), it follows that
$$
    z_j
    =
    C_j
    (\Phi_{j,k+1}x_{k+1}
    +
    \sum_{s=k+1}^{j-1}
    \Phi_{j,s+1}B_s w_s)
    +
    D_j w_j.
$$
Therefore, since
$w_{k+1},\ldots, w_j$ are independent of $x_k$ and $w_k$, then
\begin{align*}
    \cov(z_j,z_k)
    & =
    C_j \Phi_{j,k+1}
    \cov(x_{k+1},z_k)\\
    & =
    C_j \Phi_{j,k+1}
    (A_kP_k C_k^{\rT} + B_k D_k^{\rT}).
\end{align*}
Hence, by recalling the observability gramian from (\ref{PQ_def}), it follows that
\begin{equation}
\label{zz}
    \sum_{j=k+1}^{N}
    \|\cov(z_j,z_k)\|_2^2
    =
    \|\sqrt{Q_{k+1}}(A_kP_k C_k^{\rT} + B_k D_k^{\rT})\|_2^2.
\end{equation}
Now, the system $F$ is outer on the time interval $[0,N]$ if and only if $\cov(z_j,z_k) = \delta_{jk}I_r$ for all $N\> j \> k \> 0$, with $\delta_{jk}$ the Kronecker delta. In view of (\ref{covzk}) and (\ref{zz}), the outerness is equivalent to (\ref{cond}).
\endproof
\end{pf}

\end{document}